\documentclass[12pt]{JHEP3}
\usepackage{epsfig,amsmath,amssymb}

\preprint{KIAS-P08081}

\title{ Dark matter and sub-GeV hidden U(1) in GMSB models}

\author{Eung Jin Chun\footnote{ejchun@kias.re.kr} and Jong-Chul Park\footnote{jcpark@kias.re.kr}
\\Korea Institute for Advanced Study, Heogiro 87, Dongdaemun-gu, Seoul 130-722, Korea}

\abstract{Motivated by the recent PAMELA and ATIC data, one is led
to a scenario with heavy vector-like dark matter in association
with a hidden $U(1)_X$ sector below GeV scale. Realizing this idea
in the context of gauge mediated supersymmetry breaking (GMSB), a
heavy scalar component charged under $U(1)_X$ is found to be a
good dark matter candidate which can be searched for direct
scattering mediated by the Higgs boson and/or by the hidden gauge
boson.  The latter turns out to put a stringent bound on the
kinetic mixing parameter between $U(1)_X$ and $U(1)_Y$: $\theta
\lesssim 10^{-6}$. For the typical range of model parameters, we
find that the decay rates of the ordinary lightest neutralino into
hidden gauge boson/gaugino and photon/gravitino are comparable,
and the former decay mode leaves displaced vertices of lepton
pairs and missing energy with distinctive length scale larger than
20 cm for invariant lepton pair mass below 0.5 GeV. An
unsatisfactory aspect of our model is that the Sommerfeld effect
cannot raise the galactic dark matter annihilation by more than 60
times for the dark matter mass below TeV.}


\maketitle

\begin{document}

\section{Introduction}

It is now firmly established that 23\% of the energy density of
the universe consists of an unknown particle called dark matter.
Discovering the nature of dark matter would be one of the most
important tasks in current and future theoretical and experimental
investigations.  Numerous searches for galactic dark matter have
been made to observe direct signals of dark matter scattering with
nuclei and indirect evidences of dark matter annihilation to
various Standard Model particles. Recently, a number of
stimulating results toward indirect signals have been announced.

The Payload for Antimatter Matter Exploration and Light-nuclei
Astrophysics (PAMELA) collaboration has reported an excess in the
positron fraction, $e^+/(e^++e^-)$, but no excess in anti-proton
fraction $\bar{p}/p$~\cite{PAMELA}. The observed spectrum departed
from the background calculation of the cosmic-ray secondary
positron spectrum~\cite{e+Background} for energies 10--100 GeV.
This recent result is comparable with less certain excesses
observed in HEAT~\cite{HEAT} and AMS-01~\cite{AMS-01}. The
PPB-BETS balloon experiment and the Advanced Thin Ionization
Calorimeter (ATIC) instrument have also reported an excess in the
$e^++e^-$ energy spectrum of $300-800$ GeV~\cite{PPB-BETS, ATIC}.

The origin of such excesses could be astronomical objects such as
a pulsar which is expected to emit energetic electron-positron
pairs~\cite{Pulsars,GemingaPulsar,Hooper11}. On the other hand,
excessive antiparticle fluxes in galactic cosmic rays have been
considered as a primary way for indirect detection of dark matter.
It is encouraging that galactic dark matter annihilation can
easily reproduce the amplitudes and spectral shapes of the PAMELA
and ATIC data. However, there appear two puzzling aspects. First,
the dark matter annihilation cross-section  is required to be
enhanced by order of $100-1000$ compared with what is allowed by
the standard thermal relic abundance analysis.
According to a recent analysis based on $\Lambda$CDM $N$-body
simulations, however, the boost factor from clumpy matter
distribution can hardly larger than about
10~\cite{BfactorLimit1_Nsimulation}. Second, no distinctive excess
of antiprotons in the PAMELA data disfavors most conventional dark
matter candidates. A new analysis of the $\overline{p}/p$ ratio
compared to the PAMELA data puts stringent limits on possible
enhancements of the $\overline{p}/p$ flux from dark matter
annihilation~\cite{BfactorLimit2_AntiprotonAnaysis}. Many ideas
and models of dark matter annihilation or decay have already been
suggested for the explanation of the recent
observations~\cite{Neutralino, SUSYDM1, MinimalDM, ModelIndep,
DecayingU(1)x, ModelIndep2, ModelIndep3_Sommerfeld, TwoDM,
ArkaniHamed:DM, SecludedDM, TwoDM2_1metastable, Non-MinimalDM,
NWeinerLightBoson, AxionPortal, 4fermi-interaction,
Stueckelberg:PAMELA, LeptonicallyDecay, SUSYDM2, singletUED,
LeptophilicDM, U(1)x_U(1)B-L, DecayingDM, U(1)_Lmu-Ltau,
ATIC_PAMELA, ATIC_PAMELA_DecayingDM, Multi-Zurek}.
However, it appears to be a hard task to understand the nature of
enhanced hadrophobic annihilation in a consistent framework with
thermal dark matter.

\medskip

In this paper, we elaborate the idea of dark matter charged under
an extra $U(1)_X$ gauge group suggested in
Refs.~\cite{ArkaniHamed:DM, SecludedWIMP, ArkaniHamed:LHC}. The
presence of $U(1)_X$ broken at the sub-GeV scale is motivated to
explain the two intriguing features of the PAMELA data. First, the
DM annihilation to extra light gauge bosons can be enhanced
significantly by non-perturbative Sommerfeld effect
\cite{Sommerfeld-DM, Sommerfeld-Hisano, Sommerfeld-Profumo,
Sommerfeld-Cirelli, Sommerfeld-HiggsPortal}. Second, the
hadrophobic nature of the $U(1)_X$ gauge boson decay can be
understood by kinematics when its mass is below the GeV
scale~\cite{ArkaniHamed:DM, NWeinerLightBoson}. We will consider
the extra $U(1)_X$ gauge boson hidden from  the Standard Model
sector except for a small kinetic mixing  between $U(1)_X$ and
$U(1)_Y$, through which the $U(1)_X$ gauge bosons can decay mostly
into lepton-antileptons. It was shown long ago that kinetic mixing
can exist between  two $U(1)$ gauge bosons without violating the
gauge-invariance and renormalizability \cite{KineticMixing}. The
mixing between an unbroken extra $U(1)$ and $U(1)_{em}$ has been
used to explain other anomalous astronomical observation such as
galactic 511 keV $\gamma$-rays~\cite{511millicharged}.

Breaking $U(1)_X$ below the GeV scale can arise naturally by
radiative mechanism, in particular, in the context of gauge
mediated supersymmetry breaking (GMSB) \cite{ArkaniHamed:LHC}.  We
will first present a concrete way of realizing such a scheme,
which predicts a $U(1)_X$ charged scalar field as a dark matter
candidate. The dark matter mass in the range of 600--1000 GeV is
preferred for the simultaneous explanation of the PAMELA and ATIC
data. Unfortunately, the Sommerfeld enhancement factor of our
scenario is found to be about 40--60 for this mass range. The
assumption of some clumpy distribution of dark matter would be a
reasonable additional source for further increase of the boost
factor. Our dark matter candidate can yield observable signals for
direct detection through elastic scatterings mediated by the
hidden gauge boson $X$ or Higgs bosons. In the first case, we draw
a severe constraint on the kinetic mixing parameter:
$\theta\lesssim 10^{-6}$ for $m_X\sim 0.4$ GeV.  One of the
interesting consequences of our scenario is that the ordinary
lightest supersymmetric particle (OLSP), typically neutralino,
decays instantaneously to the $X$ boson and its superpartner
$\tilde{X}$ through one-loop diagram with heavy dark matter
superfields in the loop.  Then, the produced $X$ boson
subsequently decays to two leptons with a decay length typically
larger than 10 cm, which may be observed at the LHC.

\section{Spontaneous breaking of hidden $U(1)_X$ by gauge mediation}

Let us start with working in gauge mediated supersymmetry breaking
(GMSB) models \cite{gmsb} as a promising way of realizing the
sub-GeV $U(1)_X$ sector.  In such a scheme, the origin of a heavy
mass of dark matter from the hidden $U(1)_X$ sector and  the $\mu$
term can be related in the context of non-minimal supersymmetric
standard model \cite{ArkaniHamed:LHC}:
 \begin{equation} \label{eq:nmssm}
 W= {\lambda_S \over 3} S^3 + \lambda_q S q^\prime q^{\prime c} +
  \lambda_H S H_1 H_2 + \lambda_\Psi S \Psi \Psi^c\;,
 \end{equation}
where ($\Psi, \Psi^c$) is the Dirac pair carrying $U(1)_X$ charge
($+1, -1$). Note that the second term with extra quark pairs is
introduced to generate sufficiently large negative mass-squared
for $S$ and thus a large vacuum expectation value $v_S$, which
will lead to proper electroweak symmetry breaking
\cite{agashe97}.  A similar mechanism will be used to break
$U(1)_X$ at the sub-GeV scale in the below. Having generated $v_S$,
one obtains $\mu$ and $B$ terms for the Higgs fields:
$\mu=\lambda_H v_S$ and $B=\lambda_S v_S$.  Similarly, we also
obtain the supersymmetric mass of hidden Dirac fermions $m_\psi
\equiv \lambda_\Psi v_S$ and the corresponding (Dirac) soft
bilinear term:
 \begin{equation}
  V_{soft} = B m_\psi \tilde{\psi} \tilde{\psi}^c + h.c. \,.
 \end{equation}
Because of this  mixing mass term, two hidden scalars
$\tilde{\psi}$ and $\tilde{\psi}^c$ have mass splitting  and the
mass eigenstates denoted by $\tilde{\psi}_{1,2}$ have the masses
$m_{\tilde{\psi}_{1,2}}$ where $m_{\tilde{\psi}_{1,2}}^2= m_\psi^2
\mp B m_\psi$ taking $B$ is real positive.  Therefore, the lighter
scalar $\tilde{\psi}_1$, being lighter than the Dirac fermion
($\psi,\psi^c$), will be  our dark matter particle.

Let us turn to supersymmetry breaking fed into the hidden $U(1)_X$
sector.  Note that the Dirac
superfield ($\Psi,\Psi^c$) plays a role of `messenger field' for
the hidden sector supersymmetry breaking whose size is controlled
by the $B$ parameter. That is, the supersymmetry breaking masses
of the hidden gaugino $\tilde{X}$ and the scalar component
$\phi_x$ of a hidden superfield $\Phi_x$, carrying $U(1)_X$ charge
$x$, are given by
 \begin{eqnarray}
 m_{\tilde{X}} &=& {\alpha_X \over 4\pi} B \nonumber\\
 m^2_{\phi_x} &=&  2 x^2 m_{\tilde{X}}^2
 \end{eqnarray}
which are defined at the `messenger scale' $m_\psi$. Here $\alpha_X = g_X^2/4\pi$
is the $U(1)_X$ gauge fine structure constant.
The reference value for the hidden gaugino mass is $m_{\tilde{X}} \approx 0.3$ GeV
for $\alpha_X=\alpha$ and $B=500$ GeV.
Consider now the following hidden sector superpotential:
 \begin{equation}
  W = \lambda_1 \Phi_0 \Phi_{z} \Phi_{-z} + \lambda_p \Phi_0 \Phi_{pz} \Phi_{-pz}
  + {\lambda_0\over 3} \Phi_0^3 \,,
 \end{equation}
where the subscripts denote the $U(1)_X$ charges.
Denoting $m_x$ as the soft mass of the scalar component $\phi_x$, we get
$m_{pz}^2 = p^2 m_z^2$ and $m_0^2=0$ at the messenger scale $m_\psi$.  Then, one can
generate a large negative mass-squared for $\phi_0$ down around the scale $m_{pz}$
assuming   $m_{pz}^2\gg m^2_z$ ($p>1$) and the Yukawa coupling $\lambda_{p}$ of order one:
 \begin{equation} \label{m0square}
 m_0^2 \approx - n {p^2 \lambda_{p}^2\over 4\pi^2} m_z^2 \ln {m_\psi\over
 p m_z}\;,
 \end{equation}
where $n$ is the number of the $\Phi_{\pm pz}$ pairs. This will
induce a sizable mixing  vacuum expectation value $v_0$ of
$\phi_0$ determined to be $v_0^2=-m_0^2/2\lambda_0^2$. Then the
scalar potential for $\phi_{\pm z}$ is given by
 \begin{align}
 V = \lambda_1^2 \left| \phi_z \phi_{-z}\right|^2 + \left[ \lambda_1\lambda_0 v_0^2 \phi_z\phi_{-z}
 + c.c.\right] + (\lambda_1^2 v_0^2 + m_z^2) [|\phi_z|^2+|\phi_{-z}|^2] \,.
 \end{align}
If the mixing mass term of $\phi_z\phi_{-z}$ is larger than the
diagonal mass $m_z^2+\lambda_1^2 v_0^2$, there appears a direction
of negative mass-squared and thus $U(1)_X$ symmetry breaking
occurs.  For the minimization of the above potential, we examine
the D-flat direction $\phi_{-z} = \phi_z$ which has the mass
eigenvalues $m^2_{\pm} = 2(m_z^2 + \lambda_1^2 v_0^2 \pm \lambda_1
\lambda_0 v_0^2)$. To get $m^2_-<0$ and thus $v_z^2 \equiv \langle
\phi_z\rangle^2 = - m^2_0/\lambda_1^2$, one needs to have
 \begin{equation}
  -m_0^2 > 8 m_z^2
  \end{equation}
assuming $\lambda_0=2\lambda_1 < \lambda_p \sim 1$.
This condition can be generically met
by adjusting the parameters $n,p$ in Eq.~(\ref{m0square}).
Given $v_z$, the $X$ gauge boson
gets the mass
 \begin{equation}
 m_X = 2 g_X z v_z \,.
 \end{equation}
The $X$ boson mass is therefore settled down around the hidden
gaugino mass scale $m_{\tilde{X}}$. However, their specific values
will depend on the choice of the parameters $n,p$ and the
couplings $\lambda_{0,1,p}$.

Note that the hidden $U(1)_X$ sector can have a communication with
the Standard Model sector through the Yukawa couplings,
$\lambda_H$ and $\lambda_\Psi$. In particular, our scalar dark
matter $\tilde{\psi}_1$ can annihilate to quarks and leptons
through the Higgs contact interaction:
\begin{align} \label{HHDM}
{\cal L} = -{1\over2} \lambda_H\lambda_\Psi H_1H_2 \widetilde{\psi}_1\widetilde{\psi}_1^* + h.c. \,.
\end{align}
We want this channel to be
subdominant to the dark matter annihilation into
the light hidden fields as alluded in the
introduction.  For this purpose, we will assume
 \begin{equation} \label{HHban}
 \lambda_H\lambda_\Psi \ll g_X^2
 \end{equation}
throughout this paper.  Then, there can also be kinetic mixing
between $U(1)_X$ and $U(1)_Y$ through which the $X$ gauge boson can
decay to leptons as will be discussed in the next section.
The interaction (\ref{HHDM})
will give an important contribution to dark matter scattering
off nuclei through the Higgs exchange.  In the decoupling limit of
heavy neutral Higgs, the interaction between $\widetilde{\psi}_1$
and the light Higgs $h^0$ is given by
\begin{align}
{\cal L} =
 \lambda v h^0 \widetilde{\psi}_1\widetilde{\psi}_1^*\,,
 \label{hDMDM}
\end{align}
where $\lambda\equiv s_\beta c_\beta \lambda_H \lambda_\Psi/2$.
Here $\langle H_{1,2}^0\rangle = v_{1,2}/\sqrt{2}$ and
$\tan\beta\equiv v_2/v_1$.

\section{Kinetic mixing between gauge bosons/gauginos of
$SU(2)_{\rm L}\times U(1)_{\rm Y}$ and a hidden $U(1)_X$}

If $U(1)_X$ was unbroken, it mixes with $U(1)_{em}$ and various
interesting phenomena of milli-charged dark matter particles can
occur [See~\cite{511millicharged} and references therein, and
see~\cite{Stueckelberg} and references therein for the
Stueckelberg extension with kinetic mixing]. When $U(1)_X$ is
broken, the massive $X$ gauge boson mixes essentially with the $Z$
boson and leads to phenomenological consequences drastically
different from the pervious case.
Consider the gauge kinetic terms of
 $SU(2)_{\rm L}\times U(1)_{\rm Y} \times U(1)_X$ including the kinetic mixing term:
\begin{equation}
{\cal L}_{\rm
mixing}=
-\frac{\sin\theta}{2} \hat{B}^{\mu\nu}\hat{X}_{\mu\nu} \,.
\end{equation}
The canonical form of the gauge kinetic term can be  made
by the following transformation:
\begin{eqnarray} \label{canon-kin}
\left(
  \begin{array}{c}
    \hat{B}_\mu \\
    \hat{X}_\mu \\
  \end{array}
\right) = \left(
            \begin{array}{cc}
              \sec\theta & 0 \\
              -\tan\theta & 1 \\
            \end{array}
          \right) \left(
                    \begin{array}{c}
                      B'_\mu \\
                      X'_\mu \\
                    \end{array}
                  \right)\,.
\end{eqnarray}
After the $SU(2)_L\times U(1)_Y \times U(1)_X$ breaking, the
canonical gauge fields get masses and mixing.
Diagonalizing the gauge boson mass matrix,  one finds the
relation between the original gauge eigenstates $\hat{W}^3,
\hat{B}$, and $\hat{X}$ and mass eigenstates $Z, A$, and X:
\begin{align}
\left(
  \begin{array}{c}
    \hat{W}^3_\mu \\
    \hat{B}_\mu \\
    \hat{X}_\mu\\
  \end{array}
\right) = \left(\begin{array}{ccc}
 c_W & s_W & -\frac{c_W s_W m_Z^2}{m_Z^2-m_X^2}\; \theta \\
 -s_W & c_W & -\frac{c_W^2 m_Z^2-m_X^2}{m_Z^2-m_X^2}\;\theta \\
 \frac{s_W m_Z^2}{m_Z^2-m_X^2}\; \theta & 0 & 1\\
\end{array}\right) \left(
                    \begin{array}{c}
                      Z_\mu \\
                      A_\mu \\
                      X_\mu \\
                    \end{array}
                  \right)
+\mathcal{O}(\theta^2) \,. \label{basischange}
\end{align}
Let us now present some couplings relevant for our discussions in
the limit of $\theta \ll1$ and $m_X \ll m_Z$. The scalar dark
matter $\widetilde{\psi}_1$ has the gauge interactions with $X$
and $Z$ as follows:
\begin{align}
{\cal L} = i g_X
(\widetilde{\psi}_1^*\partial^\mu\widetilde{\psi}_1
-\partial^\mu\widetilde{\psi}_1^*\widetilde{\psi}_1) (X_\mu +
\theta s_W Z_{\mu}) +g_X^2 \widetilde{\psi}_1^* \widetilde{\psi}_1
X^\mu X_\mu\;.
\label{hidden-shift}
\end{align}
The hidden gauge boson $X$ couples also to the visible sector
particles, which is described by
\begin{align}
\label{Xff}
&{\cal L}= \theta g^\prime X_\mu \, \overline{f}\gamma^\mu\Gamma_{fX}f \\
{\rm with} \quad &\Gamma_{fX}\approx - Q_f
c_W^2\;\;\mbox{and}\;\;\; \Gamma_{\nu X}=-\frac{m_X^2}{2
m_Z^2}P_L\,,
 \nonumber
\end{align}
where $Q_f$ is the electromagnetic charge of a quark or lepton and
again the sub-leading terms of order of $m_X^2/m_Z^2$ are
neglected except for neutrinos. In addition, the coupling of
$XWW$, $g_{XWW}$, is given by $g_{XWW}= -\theta s_W g_{ZWW}$ where
$g_{ZWW}$ is the SM coupling of $ZWW$.

The supersymmetric counterpart of Eq.~(\ref{canon-kin}) can be
read straightforwardly~\cite{SUSY-HiddenU1}, that is, the original
hidden gaugino can be replaced by the relation,
 \begin{equation}
 \hat{\tilde{X}} =  \tilde{X} -\theta \tilde{B}
 \end{equation}
leading to the bino interaction to the hidden sector fermion and
scalar. In the right-hand side of the above equation, the gauginos
can be considered as the mass eigenstates as the mixing mass
$\theta m_{\tilde{X}}$ between $\tilde{X}$ and $\tilde{B}$ arising
from the previous consideration  can be safely neglected in our
discussion. The coupling of bino to light hidden sector bosons and
fermions given by
 \begin{equation} \label{Bphiphi}
  {\cal L} = \theta g_X \tilde{B} \tilde{\phi} \phi^* + h.c.
 \end{equation}
provides a decay channel of the ordinary lightest supersymmetric
particle in the visible sector.

\section{Relic density, positron/electron excess and direct detection}

With the assumption of Eq.~(\ref{HHban}) and
$\theta \ll 1$, which will be discussed shortly, we can
neglect the dark matter annihilation to the Standard Model
particles through Higgs and $Z,X$ gauge boson channels, namely,
$\widetilde{\psi}_1\widetilde{\psi}_1^* \rightarrow HH,  XZ$ and
$ZZ$.  Then, the
dark matter relic density is determined by the process
$\widetilde{\psi}_1\widetilde{\psi}_1^* \rightarrow XX$ having the
cross-section:
\begin{align}
\langle \sigma v\rangle_{XX}
\simeq \frac{4\pi \alpha^2_X}{m_{\widetilde{\psi}_1}^2} \,.
\end{align}
The present relic density of $\widetilde{\psi}_1$ is then given by
\begin{align}
\Omega_{\widetilde{\psi}_1} h^2 \simeq\;& \frac{1.07\times10^9\;
{\rm GeV}^{-1}}{M_{pl}} \frac{x_F}{\sqrt{g_*}} \frac{1}{\langle
\sigma v\rangle_{XX} }\nonumber\\
\approx\;& \frac{2.17\times10^{-10}\;{\rm GeV}^{-2}}{\langle
\sigma v\rangle_{XX}  }\;, \label{Omegav}
\end{align}
where $g_*$ counts the number of relativistic degrees of freedom
and $x_F\equiv m/T_F$~\cite{Bertone:2004pz}. The observed value of
$\Omega_{\rm DM}h^2 \simeq 0.1143$~\cite{WMAP5} fixes the hidden
gauge coupling constant in terms of the dark matter mass:
\begin{equation}
\alpha_X \simeq 1.58\, \alpha  \left( {m}_{\widetilde{\psi}_1}
 \over \mbox{TeV} \right) \;.\label{mass-eta}
\end{equation}
%

\begin{figure}[t]
\begin{center}
\includegraphics[width=0.80\linewidth]{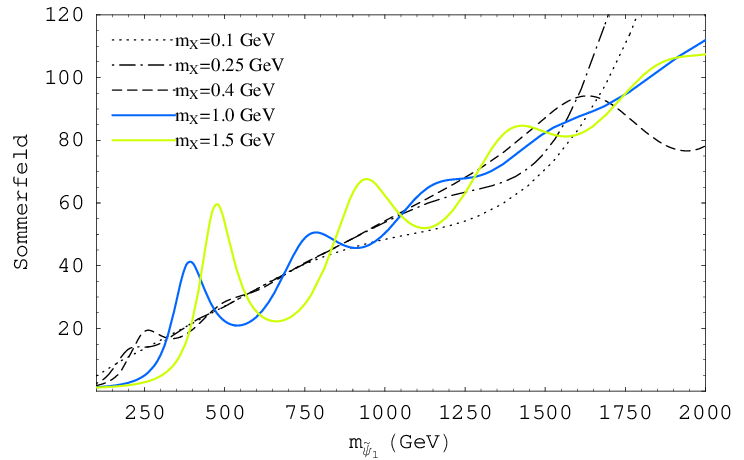}
\end{center}
\caption{The Sommerfeld enhancement factor $S$ as a function of
$m_{\widetilde{\psi}_1}$ for various $m_X$.}\label{Sommerfeld}
\end{figure}

One of the troubles for the thermal dark matter explaining the recent positron/electron
excesses from the galactic dark matter annihilation is the big gap
between the cross section $\langle \sigma v\rangle_{\rm GAL}$
required by the PAMELA and ATIC results and the cross section
$\langle \sigma v\rangle_{\rm F.O.}$ required by the dark matter relic density
at the epoch of freeze-out. Their ratio is shown to be
\begin{align}
\frac{\langle \sigma v\rangle_{\rm GAL}}{\langle \sigma v\rangle_{\rm F.O.}}
\sim 10^{2-3} \left(\frac{M}{\rm TeV}\right)^2=B_e S\;,
\label{Enhancement}
\end{align}
where $B_{e}$ is the boost factor for the positron/electron and $S$ is
the Sommerfeld enhancement factor~\cite{ModelIndep3_Sommerfeld}.
In our model,  the Sommerfeld enhancement  arises due to the light
gauge boson, and is calculated as $S=|\tilde{\psi}_1(\infty)/\tilde{\psi}_1(0)|^2$
after solving the following equation:
\begin{equation}
 -{1\over m_{\tilde{\psi}_1} } {d^2 \tilde{\psi}_1(r) \over d r^2 }
 -{\alpha_X \over r} e^{-m_X r}\tilde{\psi}_1(r) =  m_{\tilde{\psi}_1} \beta^2 \tilde{\psi}_1(r)
\end{equation}
with the out-going boundary condition,
$\tilde{\psi}'_1(\infty)/\tilde{\psi}_1(\infty)=i
m_{\tilde{\psi}_1} \beta$ where $\beta$ is the dark matter
velocity \cite{ModelIndep3_Sommerfeld, ArkaniHamed:DM,
Sommerfeld-Hisano, Sommerfeld-Profumo, Sommerfeld-Cirelli,
Sommerfeld-HiggsPortal}. Requiring the relic density condition
(\ref{mass-eta}), the Sommerfeld factor becomes a function of two
input parameters, $m_{\widetilde{\psi}_1}$ and $m_X$, which is
shown in Fig.~\ref{Sommerfeld}. As can be seen, the Sommerfeld
enhancement factor is around 50 for $m_{\widetilde{\psi}_1}\sim$
800 GeV, and increases almost linearly as $m_{\widetilde{\psi}_1}$
increases except resonance effect for certain values.  Our
enhancement factor turns out to be insufficient to explain the
PAMELA/ATIC data. This may indicate the presence of the combined
effect with the boost factor $\lesssim 10$ from clumpy dark matter
distribution \cite{BfactorLimit1_Nsimulation}.



\medskip

\begin{figure}[t]
\begin{center}
\includegraphics[width=0.80\linewidth]{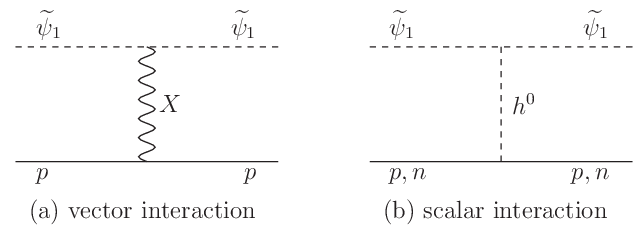}
\end{center}
\caption{Diagrams relevant to $\widetilde{\psi}_1$-nucleon elastic
scattering.}\label{detection}
\end{figure}

Given the interactions (\ref{hDMDM},\ref{hidden-shift},\ref{Xff})
of the hidden matter to the visible particles, some of our
parameter space is constrained by dark matter experiments for
direct detection. First, the dark matter  $\widetilde{\psi}_1$
has the effective vector interactions with quarks due to their
couplings to the $X$ gauge boson:
\begin{align}
\mathcal{L}_{\rm vector}^q =b_q i(\partial^\mu\widetilde{\psi}_1^*
\widetilde{\psi}_1 -\widetilde{\psi}_1^*
\partial^\mu\widetilde{\psi}_1) \;\overline{q}\gamma^\mu q\;,
\label{b-q}
\end{align}
where $b_q \equiv \theta g_X g^\prime c_W^2 Q_q/m_X^2$. Now that
the vector current is conserved, the contributions of each quark
in a nucleus add coherently. In addition, sea quarks and gluons
cannot contribute to the vector current. Thus, the
$\widetilde{\psi}_1$-proton/neutron interactions can be expressed
with the replaced couplings, $b_p=2b_u+b_d$ and $b_n=b_u+2b_d$.
Note that we have $b_n\approx0$ as $X$ couples to the
electromagnetic charge in the leading order. As a result, the
$\tilde{\psi}_1$-nucleus interaction is given by
\begin{equation}
\mathcal{L}_{\rm vector}^N
=b_N i(\partial^\mu\widetilde{\psi}_1^* \widetilde{\psi}_1
-\widetilde{\psi}_1^* \partial^\mu\widetilde{\psi}_1)\;
\overline{N}\gamma^\mu N\;,
\end{equation}
where $b_N=Zb_p+(A-Z)b_n = \theta g_X g^\prime c_W^2Z/m_X^2$ and the nucleus $N$ has
the atomic number $Z$ and the atomic weight $A$.
Therefore, the standard total cross section for the
$\widetilde{\psi}_1$-nucleus vector interaction is given
by~\cite{Jungman:1995df}
\begin{align}
\sigma_{\rm vector}^{\widetilde{\psi}_1-N}
&= \frac{m_{\widetilde{\psi}_1}^2m_N^2b_N^2}
{64\pi(m_{\widetilde{\psi}_1}+m_N)^2}\nonumber\\
&\simeq 1.25\times 10^{4}\frac{\overline{m}_{\widetilde{\psi}_1}^2
\overline{m}_N^2}
{(\overline{m}_{\widetilde{\psi}_1}+\overline{m}_N)^2}
\theta^2 {\alpha_X\over \alpha} \frac{Z^2}{\overline{m}_X^4}\;{\rm pb}\;,
\end{align}
where $\overline{m}_i\equiv m_i/$GeV. Recall that the vector
interaction leads only to spin-independent cross-section. To
compare with experiments, we plot the $\widetilde{\psi}_1$-proton
cross section for vector interactions given by
\begin{align}
\sigma_{\rm vector}^{\widetilde{\psi}_1-p}
= \sigma_{\rm vector}^{\widetilde{\psi}_1-N}\frac{1}{Z^2}
\frac{\mu_p^2}{\mu_N^2}\;,
\end{align}
where $\mu_{p, N}$ are the reduced masses for the
$\widetilde{\psi}_1$-proton and
$\widetilde{\psi}_1$-nucleus.

In addition to the vector interaction, the scalar dark matter
$\widetilde{\psi}_1$ interacts with nucleons through $t$-channel
Higgs exchange driven by Eq.~(\ref{hDMDM}). The
$\widetilde{\psi}_1$-nucleon cross section for the scalar
interaction is adequately estimated in~\cite{ScalarDMCrossSec}:
\begin{align}
\sigma_{\rm scalar}^{\widetilde{\psi}_1-n, p} \approx\;&
\left(\frac{\lambda\;0.34\;{\rm GeV}}{m_h^2 \pi}\right)^2
\left(\frac{m_p}{m_{\widetilde{\psi}_1}+m_p}\right)^2\nonumber\\
\simeq\;& 4.56\times10^6 \frac{\lambda^2}{\overline{m}_h^4}
\frac{\overline{m}_p^2}
{(\overline{m}_{\widetilde{\psi}_1}+\overline{m}_p)^2} \;{\rm
pb}\;,
\end{align}
where $\overline{m}_i\equiv m_i/$GeV.
\begin{figure}[t]
\begin{center}
\includegraphics[width=0.80\linewidth]{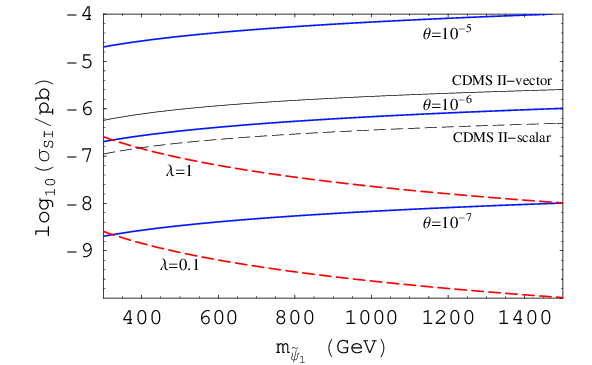}
\end{center}
\caption{Exclusion plot for the spin-independent $\tilde{\psi}_1$-nucleon cross-section
$\sigma_{\rm SI}$.
The solid lines are the cross-sections via vector interactions
with $m_X=0.4$ GeV corresponding to
$\theta=10^{-5}, 10^{-6}$, and $10^{-7}$, respectively.
The dashed lines are  through Higgs exchange
corresponding to $\lambda=1$ and 0.1, respectively, taking
$m_h=115$ GeV.
The thin solid (dashed) line shows the CDMS II limit for the vector (scalar)
interaction.}\label{DirectDetec}
\end{figure}
If the Standard Model Higgs and hidden light higgs have the contact
interaction term, $\lambda_{\phi h}\phi^2h^2$, the hidden higgs
mixes with the SM Higgs when both fields have vacuum expectation values.
Therefore, the
t-channel exchange of the light hidden scalar through this mixing
possibly dominates in some cases~\cite{LightScalarExchange}.
However, in our model, the visible and hidden sectors are
separated, and so no such contact interaction exists at the
tree-level. The t-channel light scalar exchange can
effectively arise only from three-loop diagrams, which can be safely neglected.

The Korea Invisible Mass Search (KIMS) experiment provides the
most stringent limit on the spin-dependent  interaction for a pure
proton case~\cite{KIMS}, and the Cryogenic Dark Matter Search
(CDMS II) experiment sets the strongest limit on the
spin-independent  WIMP-nucleon interaction for a WIMP mass larger
than $\sim 100$ GeV~\cite{CDMS II}. However, the latter is only
relevant in this analysis. In Fig.~\ref{DirectDetec}, we present
the  $\tilde{\psi}_1$-nucleon cross-sections via vector
interactions (solid lines) as a function of
$m_{\widetilde{\psi}_1}$ for typical kinetic mixing parameters,
$\theta=10^{-5}, 10^{-6}$, and $10^{-7}$. The  scalar interaction
cross-sections (dashed lines) are also shown as a function of
$m_{\widetilde{\psi}_1}$ for $\lambda=1$ and 0.1. The limits from
CDMS II experiment is shown as thin line respectively for the
vector and scalar interaction. The vector interaction puts a
strong bound on the kinetic mixing parameter:
\begin{equation}
\theta \leq 2\times10^{-6} \left( m_X \over 0.4 \, \mbox{GeV}
\right)^2\,,
\end{equation}
whereas the scalar interaction puts almost no bound on
$\lambda=s_\beta c_\beta \lambda_H\lambda_\Psi/2$.
We remark that in the framework of the {\it inelastic dark
matter}~\cite{IDM} one can evade this stringent bound on the
kinetic mixing parameter $\theta$ as discussed
in~\cite{ArkaniHamed:DM}, where the dark matter is part of a
multiplet of a non-Abelian gauge group and small mass splittings
exist between these states. In this case, the most stringent bound
on kinetic mixing comes from the anomalous magnetic moment of
muon: $\theta \lesssim
9\times10^{-3}(m_X/0.4\mbox{GeV})$~\cite{UbosonAMMofMuon}.

\section{Visible OLSP decay and Displaced vertices of lepton pairs}

\begin{figure}[t]
\begin{center}
\includegraphics[width=0.80\linewidth]{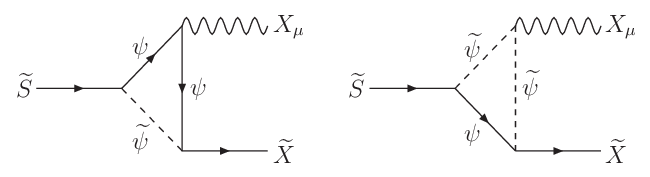}
\end{center}
\caption{One-loop diagrams responsible for the decay of singlino
$\widetilde{S}$ to $X\widetilde{X}$.}\label{Feynman}
\end{figure}

In our scenario, the ordinary lightest supersymmetric particle
(OLSP) $\chi^0_1$, which is typically a linear
combination of neutralinos including singlino $\tilde{S}$, can
decay to the hidden sector particles either through the kinetic
mixing (\ref{Bphiphi}) or through the singlino coupling to the
dark matter superfield components $\psi, \tilde{\psi}$
(\ref{eq:nmssm}). The latter leads to the OLSP decay into hidden
gauge boson $X$ and gaugino $\tilde{X}$ resulting from the
one-loop induced $\tilde{S}-X-\tilde{X}$ interaction as shown in
Fig.~\ref{Feynman}. The corresponding effective Lagrangian is
\begin{align}
& \mathcal{L} = \frac12 (C_1+C_2) \overline{\widetilde{X}}
\sigma^{\mu\nu}\widetilde{S}X_{\mu\nu}\\
{\rm with}\quad & C_i=\frac{m_\psi}{16\pi^2 m_i^2}
\frac{\lambda_\Psi g_X^2}{\sqrt{2}}J(x_i)\;,\nonumber
\end{align}
where $x_i=m_\psi^2/m_i^2$, $m_{1, 2}^2
\equiv m_{\widetilde{\psi}_{1, 2}}^2 = m_\psi^2 \mp B m_\psi$, and
$J(x)\equiv \frac{1}{(1-x)^3}[-2+2x-(1+x)\ln x]$.
In the limit of $x \rightarrow 1$, we have $J(x)=1/6$ and the decay
rate of the singlino $\tilde{S}$ becomes
\begin{align}
\Gamma(\widetilde{S}\rightarrow X\widetilde{X})=\frac{1}{8\pi}\;
m_{\widetilde{S}}^3\;(C_1+C_2)^2\approx
\frac{\lambda_\Psi^2\alpha_X^2}{2304\pi^3}\;
\frac{m_{\widetilde{S}}^3}{m_\psi^2}\;,
\end{align}
if the singlino is the OLSP.
In the case of the bino OLSP, the decay rate of the bino $\tilde{B}$
to hidden higgs and higgsinos ($\phi$ and $\widetilde{\phi}$) via the kinetic
mixing is given by
\begin{align}
\Gamma(\widetilde{B}\rightarrow
\phi\widetilde{\phi})\approx \frac{\alpha_X\theta^2}{2}\;
m_{\widetilde{B}}\;.
\end{align}
Considering the OLSP  having the  bino (singlino) component with a
fraction $c_{\widetilde{B}}$ ($c_{\widetilde{S}}$), we get the
ratio between two decay modes as
\begin{align}
\frac{c_{\widetilde{B}}^2 \Gamma(\widetilde{B}\rightarrow
\phi\widetilde{\phi})}{c_{\widetilde{S}}^2
\Gamma(\widetilde{S}\rightarrow X\widetilde{X})} \approx\;
&5\times10^{-4}
\left(\frac{c_{\widetilde{B}}}{c_{\widetilde{S}}}\right)^2
\left(\frac{0.1}{\lambda_\Psi^2}\right)\left(\frac{\theta}{10^{-6}}\right)^2 \nonumber\\
& \left(\frac{m_\psi}{700\;{\rm GeV}}\right)^2
\left(\frac{200\;{\rm GeV}}{m_{\widetilde{\chi}^0_1}}\right)^2\;,
\end{align}
where  we have used Eq.~(\ref{mass-eta}) and assumed
$m_{\widetilde{\psi}_1} \approx m_\psi$ for simplicity. Thus, we
can conclude that $\chi_1^0$ mostly decays to a hidden
gauge boson and gaugino  $X\widetilde{X}$ in a reasonable choice
of the parameter space. From the above calculations, the decay
length of $\chi_1^0$ is determined to be
\begin{align} \label{chi1X}
l_{\chi_1^0 \rightarrow X\widetilde{X}} \approx
\left(\frac{0.1}{c_{\widetilde{S}}}\right)^2
\left(\frac{0.1}{\lambda_\Psi^2}\right) \left(\frac{200\;{\rm
GeV}}{m_{\widetilde{\chi}_1^0}}\right)^3 \times 10^{-3}\; {\rm cm}
\end{align}
ensuring that the OLSP decays well inside a detector.
The decay-produced hidden gauge
boson $X$ decays back to the Standard Model fermion pair $f\overline{f}$ via
the interaction~(\ref{Xff}), and the corresponding decay length is
\begin{align}
l_{X \rightarrow f\overline{f}} \approx
\left(\frac{1}{Q_f}\right)^2 \left(\frac{10^{-6}}{\theta}\right)^2
\left(\frac{0.4\;{\rm GeV}}{m_X}\right) \times 25\; {\rm cm}\;.
\end{align}
As a result,  we will see a signal of energetic lepton pairs plus
missing energy with a displacement vertex of $\mathcal{O}(10)$ cm
as $\widetilde{X}$ can either be the LSP in the hidden sector or
decay to a hidden sector higgs and higgino.

Of course, this conclusion is valid only if the conventional decay
channel of the OLSP to the gravitino $\psi_{3/2}$ is not too much more efficient than
the above process. In GMSB models, supersymetry breaking is
driven by a hidden strong dynamics which is supposed to generate
supersymmetry breaking F-term with $\sqrt{F} \gtrsim 100$ TeV
\cite{gmsb}.  Then the OLSP decay length for the process $\chi^0_1
\to \gamma \psi_{3/2}$ is
\begin{equation} \label{chi1gravitino}
 l_{\chi^0_1 \to \gamma \psi_{3/2}} \approx \left( \sqrt{F} \over 10^5 \,\mbox{GeV} \right)^4
 \left( 200\,\mbox{GeV} \over m_{\chi^0_1} \right)^5 \times 3\times10^{-4}\,{\rm cm}\,.
\end{equation}
Remarkably, this number is comparable to (\ref{chi1X}), and there
may be a chance to confirm both the GMSB mechanism and the
presence of sub-GeV hidden U(1) by observing both signals for
(\ref{chi1X}) and (\ref{chi1gravitino}).

\section{Conclusion}

The presence of TeV dark matter associated with a hidden $U(1)_X$
gauge symmetry broken below the GeV scale has been motivated by
the recent results from PAMELA and ATIC. Elaborating this idea in
the framework of gauge mediated supersymmetry breaking, we obtain
some interesting constraints and prospects of the scenario.

 Explaining
the heaviness of dark matter in relation to the resolution of the
$\mu$ and $B\mu$ problem of GMSB models, it follows that the
vector-like scalar particle charged under $U(1)_X$ becomes a good
dark matter candidate for the gauge coupling $\alpha_X \sim
\alpha$ and the dark matter mass $\sim$ TeV. The other `hidden'
particles, namely the $U(1)_X$ gauge boson and gaugino, higgs
boson and higgsino, obtain masses below the GeV scale through a
radiative mechanism with properly chosen particle contents and
Yukawa couplings among them. The hidden sector communicates with
the visible (Standard Model) sector either by the Yukwa terms
generating the Higgsino and dark matter masses  or by the kinetic
mixing between $U(1)_X$ and $U(1)_Y$.  It turns out that the
kinetic mixing parameter $\theta$ receives a strong upper limit
$\theta \lesssim 10^{-6}(m_X/0.4 \,\mbox{GeV})^2$ coming from the
experimental data on direct detection of dark matter.

In this scenario, positive signals for direct dark matter
detection in future experiments may arise either from the vector
interaction controlled by $\theta$ or from the scalar interaction
exchanging the Higgs boson.  Another interesting consequence for
the LHC experiment is that the ordinary LSP decay leaves quite
distinctive signals of missing energy and energetic lepton pairs with
invariant mass below GeV and decay gaps larger than  about  10 cm.
Furthermore, the rates for this decay and the usual decay to
photon and gravitino, a characteristic of GMSB models, are found
to be comparable in typical parameter range of the model. Thus,
the scenario of the sub-GeV $U(1)_X$ sector in GMSB models may be
tested at the LHC by observing both signatures of lepton pairs
plus missing energy and photons plus missing energy.

However, we find that the boost factor for the galactic dark
matter annihilation required by the PAMELA and ATIC data can not
be obtained solely by the Sommerfeld effect in our scenario.
Typical Sommerfeld enhancement factor is in the range of $40-60$
for the dark matter masses of $600-1000$ GeV.  Thus, we may have
to invoke additional sources of the boost factor like the clumpy
distribution of dark matter.


\end{document}